\documentclass[12pt]{iopart}
\usepackage{iopams}
\usepackage{harvard}
\usepackage{color}
\usepackage{url}

\expandafter\let\csname equation*\endcsname\relax
\expandafter\let\csname endequation*\endcsname\relax
\usepackage{amsmath}
\usepackage{graphicx}
\usepackage[usenames,dvipsnames]{xcolor}
\usepackage{multirow}

\newcommand{\cm}{cm$^{-1}$}
\newcommand{\ai}{\textit{ab initio}}

\newcommand{\hp}{H$_3^+$}
\newcommand{\dpp}{D$_3^+$}

\begin{document}

\title[High accuracy calculations of the spectrum of H$_3^+$]{High accuracy calculations of the rotation-vibration spectrum of H$_3^+$}

\author{Jonathan Tennyson$^{1,5}$, Oleg L. Polyansky$^{1,2}$,  Nikolai F. Zobov$^2$, Alexander Alijah$^3$, and Attila G. Cs\'asz\'ar$^4$}

\address{$^1$Department of Physics \& Astronomy, University College London,
Gower St.,London, WC1E 6BT, UK\\
$^{2}$Institute of Applied Physics, Russian Academy of Sciences, Ulyanov Street 
46, Nizhny Novgorod 603950, Russia\\
$^{3}$Groupe de Spectrom\'etrie Mol\'eculaire et Atmosph\'erique,
GSMA, UMR CNRS 7331, Universit\'e de Reims Champagne-Ardenne, France\\
$^4$Institute  of Chemistry, E\"otv\"os Lor\'and University and
MTA-ELTE Complex Chemical Systems Research Group,
H-1518 Budapest 112, P.O. Box 32, Hungary}

\ead{$^5$j.tennyson@ucl.ac.uk}

\begin{abstract}
Calculation of the rotation-vibration spectrum of H$_3^+$, as well as of its
deuterated isotopologues, with near-spectroscopic accuracy requires
the development of sophisticated theoretical models, methods, and codes.  
The present paper reviews the state-of-the-art in these fields. 
Computation of rovibrational states on a given potential energy surface (PES) has 
now become standard for triatomic molecules, 
at least up to intermediate energies, due to developments 
achieved by the present authors and others.
However, highly accurate Born--Oppenheimer (BO) energies 
leading to highly accurate PESs are not accessible even for this two-electron
system using conventional electronic structure procedures 
(\textit{e.g.}, configuration-interaction	or coupled-cluster techniques
with extrapolation to the complete (atom-centred Gaussian) basis set limit). 
For this purpose highly specialized techniques must be used, \textit{e.g.}, 
those employing explicitly correlated Gaussians and nonlinear parameter optimizations.
It has also become evident that a very dense grid of \ai\ points is required 
to obtain reliable representations of the computed points extending 
from the minimum to the asymptotic limits.
Furthermore, adiabatic, relativistic, and quantum electrodynamics (QED) correction
terms need to be considered to achieve near-spectroscopic accuracy during
calculation of the rotation-vibration spectrum of H$_3^+$. 
The remaining and most intractable problem is then the treatment of the effects of 
non-adiabatic coupling on the rovibrational energies, which, in the worst cases,
may lead to corrections on the order of several \cm.
A promising way of handling this difficulty is the further development of effective, 
motion- or even coordinate-dependent, masses and mass surfaces.  
Finally, the unresolved challenge of how to describe and elucidate the experimental 
pre-dissociation spectra of H$_3^+$ and its isotopologues is discussed.
\end{abstract}

\maketitle

\section{Introduction}

The highly stable molecular ion H$_3^+$, the prototype of a three-centre two-electron
(3c-2e) chemical bond, is rapidly formed in an exoergic (exergonic) reaction 
following the collision of molecular hydrogen and its cation,
\begin{equation}
 {\rm H}_2 + {\rm H}_2^+ \rightarrow {\rm H}_3^+ +{\rm H}. 
\end{equation}
The high stability of the ion means that H$_3^+$ is the dominant molecular ion 
in cold hydrogen plasmas, which of course make up much of the known Universe. 
H$_3^+$ has long been thought to be the initiator of much of gas-phase interstellar
chemistry via the ion--molecule reactions H$_3^+$ + X $\rightarrow$ HX$^+$ + H$_2$ (where
X can be an atom or a molecule) \cite{w73,hk73,13Oka.H3+,15Millar}, but its
rather limited spectroscopic signature, discussed in the next
section, delayed its detection in the interstellar medium (ISM).

In fact, the original extra-terrestrial detection of H$_3^+$ was an {\it in situ} 
detection on Jupiter by Voyager-2 using mass spectrometry \cite{80HaGlKr.H3+}.
Coincidentally, this discovery was approximately
contemporaneous with the original laboratory measurement of the spectrum of the ion 
by \citeasnoun{T80}.  
The spectrum of H$_3^+$ was originally observed in Jupiter \cite{jt80}; 
notably, this observation relied heavily on high-accuracy 
first-principles predictions of both the frequency and the intensity of the
observed lines \cite{jt65}.  
The spectrum of H$_3^+$ has been 
observed in the atmospheres of Uranus \cite{jt127} and Saturn \cite{gj93},
but not so far of Neptune \cite{11MeStMi.H3+}.

The long campaign to detect interstellar H$_3^+$
\cite{61MaMcMe.H3+,81Okaxxx.H3+,89GeOkxx.H3+} eventually succeeded
when \citeasnoun{96GeOkxx.H3+} found the signature of H$_3^+$ in
absorption against starlight in giant molecular clouds.
\citeasnoun{98McGeHi.H3+} subsequently detected H$_3^+$ in significant
concentration in diffuse clouds using the same technique. 
Searches for H$_3^+$ elsewhere in the Universe have proved more controversial with
a tentative detection in the remnants of supernova SN1987a
\cite{jt110} and a claimed detection in a protoplanetary disk, 
which could not be verified \cite{05GoGeMc.H3+}. 
Note that while much observational work has concentrated on H$_3^+$,
models suggest that all deuterated isotopologues, even D$_3^+$, can occur
in significant quantities in the ISM \cite{04WaFlDe}.

While H$_3^+$ is a vigorous proton donor, it is often destroyed via
dissociative recombination (DR):
\begin{equation}
{\rm H}_3^+ + e \rightarrow {\rm H}_2 + {\rm H}~~~\rm{or}~~~H+H+H.
\end{equation}
Laboratory measurement of the rate of DR long proved controversial
\cite{larsson:2000,larsson:2008}, but DR is now known from
both measurement \cite{mccall:2003b,kreckel:2010} and theory
\cite{KokooulineNature:2001,santos:2007} to be rapid. 
Studies have also
shown that the DR rate is state dependent and that the high symmetry of
H$_3^+$ opens avenues for population trapping in rotationally-excited levels
\cite{jt306,jt340,jt666}, an effect which has also been observed in the
ISM \cite{02GoMcGe.H3+}.

The dynamical characteristics of the H$_3^+$ ion have been the subject 
of a number of reviews \cite{92Oka.H3+,jt122,m93,jt157,jt535,13Oka.H3+}.
The first-principles computation of the spectrum, from the microwave to the ultraviolet, 
of a molecule such as H$_3^+$ relies on a number of steps \cite{jt475} which require
extending to treat the observed spectrum above dissociation \cite{89CaMcxx.H3+}. 
The accurate prediction of energies and frequencies requires the accurate solution of the
electronic structure problem on a grid of points and the analytic
representation of these points to give a potential energy surface (PES).
An accurate treatment of the nuclear motion problem should then
follow. This is facilitated by the fact that use of exact nuclear-motion
kinetic energy operators, which has long  been a feature 
of nuclear motion calculations on H$_3^+$ \cite{csk78}, 
is straightforward. 
Determination of contributions to the PES \cite{98CsAlSc}, often neglected in more approximate
treatments, due to relativity, quantum electrodynamics (QED), and the
breakdown of the Born--Oppenheimer (BO) approximation also need to be computed. 
For transition intensities this also requires the calculation and representation 
of accurate dipole moment surfaces (DMS).
 
More than a decade ago, \citeasnoun{03MiPeSc} declared, based on a
model developed by two of the present authors \cite{jt236},
that the spectrum of H$_3^+$ was a solved theoretical problem.  
This model, which as we show  is characterized by a fortuitous cancellation of
errors, is discussed further below and provides our starting point for
high-accuracy computations of the rovibrational spectra of the H$_3^+$ system.  
Before that we provide a brief overview of the unique
spectroscopic properties of H$_3^+$ and its isotopologues. 
We then briefly consider first-principles
computations on isotopologues of diatomic hydrogen (a four-body system),
which can indeed be classed as solved problems. 
We then move to H$_3^+$ (a five-body system), 
considering in turn electronic structure computations,
nuclear motion treatments, and representation of effects beyond the
BO approximation, the latter being particularly important for this system.
The accurate computation of transition intensities is then considered
before we provide some comments on future directions and developments.

\section{Overview of the spectrum of H$_3^+$}
Our experimental knowledge about the rovibronic spectrum of H$_3^+$ is
rather limited.  The equilibrium structure of H$_3^+$ in its ground
electronic state has $D_{3h}$ point-group symmetry, due to the 3c -- 2e bonding present in the ion.  As a highly
symmetric species, H$_3^+$ does not possess a permanent dipole moment,
the usual prerequisite for pure rotational transitions.  It has been
suggested theoretically that distortions of the ion as it rotates
should lead to an observable spectrum \cite{86PaOk.H3+,jt72}; however,
this has yet to be seen.  Calculations also suggest that the
ion distorts in the presence of a strong magnetic field~\cite{MED13:9871}.
Similarly, there are no experimentally-known electronically-excited
states of H$_3^+$, despite considerable theoretical work on the
spectroscopy of the metastable first-excited $^3\Sigma_u^+$ state
\cite{SCH74:1934,AHL77:2771,WOR89:2344,PRE91:7204,FRI01:1183,SAN01:2182,CUE02:6012,CER03:2637,ALI03:163,ALI04:243003,VIE04:253,VIE05:3307,VAR05:285,ALI06:5499,ALI06:2889,MEN08:054301,ALI15:43}.

This leaves rotation-vibration transitions, which have been
observed in the infrared and visible regions, as the only
available spectroscopic handle on the ion. 
Even here it might appear that there are slim pickings, as
H$_3^+$ has two vibrational modes: a symmetric and hence
infrared-inactive stretching mode,  $\nu_1$, and a degenerate
bending mode, $\nu_2$.  
It was the $\nu_2$ mode that \citeasnoun{T80} originally detected. 
However, it transpires that \lq forbidden\rq\
stretching transitions can also be observed \cite{jt89,92XuRoGa.H3+}
and the overtones are also strong \cite{jt65,jt83,jt102,jt119,jt193}. 
This leads to a rich spectrum of rotation-vibration transitions, 
whose observation now extends to visible wavelengths \cite{kbr08,jt512}. 
These lines probe states which lie above the barrier to linearity of the molecular ion,
which, for H$_3^+$, lies at about 10~000 \cm\ \cite{mgo09}. 
Linearity is a monodromy point \cite{jt234} and above that energy
there are a number of added complications in the theoretical
treatment and understanding of the spectra; some of these are discussed below.

There are several compilations of H$_3^+$ experimental spectroscopic
data \cite{jt107,lindsay:2001a,13FuSzMa.H3+}; the most recent of these
was performed by \citeasnoun{13FuSzMa.H3+}, who employed the MARVEL
(Measured Active Rotation-Vibration Energy Levels) code
\cite{jt412,12FuCsi.method,16CzFuAr.marvel} to provide a list of
empirical energy levels which can be used to benchmark accurate
first-principles computations.  We note, in particular, that recent
experiments
\cite{12CrHoSiPe,13WuLiLi.H3+,13HoPeJe.H3+,15PeHoMaKo,16JuKoSc.H3+,17YuPeAm.H3+}
using frequency combs have provided some particularly high-accuracy
transition wavenumbers which can be used to benchmark future, further
improved theoretical treatments.

Spectra of deuterated H$_3^+$ isotopologues have also been studied
\cite{84LuAm,fmw86,90PoMc,16JuKoSc.H3+,17YuPeAm.H3+}.  
Both H$_2$D$^+$ and D$_2$H$^+$
have permanent dipole moments due to separation between the
centre-of-mass and the centre-of-charge; see \citeasnoun{17JuTpMu.H3+}
and references therein for a discussion of the observed rotational spectrum.  
\citeasnoun{13FuSzFaCs} provide empirical energy levels for
these asymmetric-top species.  
The infrared spectrum of D$_3^+$ has
also been observed \cite{80ShFaLaWi,94AmChCi.H3+}, although less is
known about the spectrum of this species than about those of the other
deuterated isotopologues.

No discussion of the spectroscopy of H$_3^+$ is complete without
consideration of the remarkable and rich near-dissociation spectra of
the system discovered by \citeasnoun{carrington:1982}.  This spectrum,
which was systematically mapped out over the following years
\cite{carrington:1984,carrington:1989,carrington:1993,00KeKiMc.H3+},
remains unassigned and still presents a major challenge to theory for
a system which only has two electrons.  Perhaps some spectroscopic
measurements on the near-dissociation spectrum of H$_3^+$ will be
performed in the near future using a well-characterised (cold) source
of ions and a multiphoton approach.  Such a project could be performed
hand-in-hand with theory; it would appear to be the best way to
understand the seemingly rich dynamics at and beyond the first
dissociation limit.

\begin{table}
\centering 
\centering{}
\caption{Summary of empirical (MARVEL) energy levels determined for H$_3^+$ 
\cite{13FuSzMa.H3+} and two of its
deuterated isotopologues \cite{13FuSzFaCs}, given for different components of the
experimental spectroscopic networks (SN) of the ions.}\label{Tab:MARVEL}
\begin{tabular}{llr}
\hline \hline
Molecule   &SN component  &number \\
\hline
H$_3^+$    &\textit{ortho}&259    \\
           &\textit{para} &393    \\
           &floating      &105    \\
                &sum           &757    \\
H$_2$D$^+$ &\textit{ortho}& 63     \\
           &\textit{para} & 46    \\
           &floating      & 14    \\
                &sum           &123    \\
D$_2$H$^+$ &\textit{ortho}& 52      \\
           &\textit{para} & 52    \\
           &floating      & 27    \\
                &sum           &131    \\
\hline \hline
\end{tabular}
\end{table}

\section{Empirical (MARVEL) rovibrational energy levels}
H$_3^ +$ drives the chemistry in many cold parts of the universe, where only
barrierless ion-molecule reactions are feasible, and it is also a tracer 
of the chemistry of the interstellar medium \cite{jt489}.
Therefore, it is important to know as much and as accurately as possible 
about the rovibrational energy level structure of this highly stable molecular ion.

Table~\ref{Tab:MARVEL} gives a summary of the MARVEL analyses of the
experimental spectroscopic data available for H$_3^+$, H$_2$D$^+$, and
D$_2$H$^+$.  The MARVEL analysis starts with the representation of the
observed transitions data by an experimental spectroscopic network
(SN) \cite{11CsFu,11FuCs}.  In the absence of transitions 
between \textit{ortho} and \textit{para} states, the spectroscopic
data of H$_3^+$ should form two principal components (PC) comprising
transitions within the \textit{ortho} and \textit{para} states,
respectively.  However, in practice the experimental data often define
further components not attached to the PCs of the experimental SN by
any measured transitions.  These components are known as floating
components (FC) \cite{11CsFu}.

As shown by \citeasnoun{13FuSzMa.H3+}, of the 1610 measured
transitions for H$_3^ +$ available then, reported in 26 sources, 1410
could be validated, a further 105  transitions belong to FCs, 
while the rest had to be excluded from the final MARVEL analysis.  
Despite the difficulties measuring
the spectra of an ion without a dipole moment, the spectral range
covered by the experiments is wide, between 7 and 16~506 \cm.  
This experimental dataset defines 13 vibrational band origins (VBO), with
the highest $J$ value of only 12, with a typical uncertainty of about
0.005 \cm.
Since 2013 results from two
high-accuracy measurements have been reported \cite{15PeHoMaKo,16JuKoSc.H3+}.  
These transitions improve the
accuracy of the empirically-determined energy levels but do not alter
in any significant way the conclusions detailed here.  
As of today, the
number of validated and thus recommended experimental-quality
rovibrational energy levels of H$_3^+$ is 652, of which 259 belong to
\textit{ortho}-H$_3^ +$ ($I=3/2$) and 393 to \textit{para}-H$_3^+$ ($I=1/2$),
where the quantum number $I$ refers to the total nuclear spin of the system.

There have been lot fewer experimental studies dealing with the spectra of the
deuterated ions.
In fact, 13 and 9 sources dealt with H$_2$D$^+$ and D$_2$H$^+$, respectively.
These measurements define only 7 and 6 VBOs for H$_2$D$^+$ and D$_2$H$^+$, respectively.
The scarcity of experimental data means that our understanding of the rotational states
of the (0~1~0) VBO of H$_2$D$^+$ is complete only to $J=2$,
and for all higher-lying VBOs and $J$ values the information is highly incomplete.
Two pure rotational transitions of H$_2$D$^+$ are
important from an astrochemical point of view,
the 372.4 GHz $(1_{10}-1_{11})$ transition \cite{05AmHi} of \textit{ortho}-H$_2$D$^+$
and the
1370 GHz $(1_{01}-0_{00})$ transition \cite{08AsRiMuWi} of \textit{para}-H$_2$D$^+$.
Recently \citeasnoun{17JuTpMu.H3+} measured these and related
transitions of  D$_2$H$^+$  with outstanding accuracy. 
\citeasnoun{17YuPeAm.H3+} reported
the measurement of further lines of D$_2$H$^+$  with similar accuracy.
The extensive variational computations of the rovibrational energy levels
of D$_2$H$^+$, performed by \citeasnoun{13FuSzFaCs} and  \citeasnoun{96AlBe},
are in good agreement with each other.
A reliable labeling of most of the computed rovibrational states is provided by
\citeasnoun{13FuSzFaCs}, based on a rigid rotor decomposition (RRD) 
analysis \cite{10MaFaSz.method}.
The few discrepancies between the two studies are discussed by \citeasnoun{13FuSzFaCs}.
Note that the work of \citeasnoun{13FuSzFaCs} attempted to label several measured
transitions left unlabeled by \citeasnoun{90PoMc} and
\citeasnoun{82Shy}; validation of these assignments awaits further experimental studies.

\section{Computations on diatomic hydrogen: H$_2$ and H$_2^+$}                   
Computation of the rovibrational energy levels of the H$_2$ and
H$_2^+$ molecules provides precise and accurate
information which can be used to understand similar computations on \hp.  
Although comparison with experimental results
gives the ultimate criteria for establishing the accuracy of a
theoretical model, it does not easily provide estimates of the accuracy of
the components of the model.
In particular, the separate comparison of the BO
energy and the adiabatic, non-adiabatic, relativistic, and
QED energy corrections cannot be executed by a direct
comparison of computed \hp\  spectra with experimental ones.  
Since the computation of energies and energy corrections can be performed almost
exactly for H$_2$ and H$_2^+$, a comparison with the corresponding
values of \hp\ can give valuable insight.  
First of all, when there is
no way to even estimate the order of magnitude of certain corrections for triatomic
\hp, data on diatomic H$_2$ and H$_2^+$ do give these.  
Second, the methods of computation of energy corrections for polyatomics
often follow the methods developed for the diatomic hydrogen species.

As mentioned by \citeasnoun{Korobov06} (see also references given there), 
variational determinations of non-relativistic energies for
H$_2^+$ and HD$^+$ have reached a precision of 10$^{-15}$ --
10$^{-30}$ $E_{\rm h}$ (that is 10$^{-10}$-- 10$^{-25}$ \cm).  
This unprecedented precision is due to the simplicity of these three-particle systems.  
Note that there is a demand for such high-accuracy computations
for the purpose of high-accuracy determination of the electron--proton
mass ratio using spectroscopic techniques
\cite{16UbBaSa.H2,16UbKoEi.H2}.
Improvement of non-relativistic energies by
\citeasnoun{Moss1993} to 10$^{-5}$ \cm\   is already sufficient for
the purpose of deriving effective vibrational masses for use within
the \citeasnoun{80BuMoxx} non-adiabatic model.  
This model was an important development, as it could be validated using highly accurate
experimental H$_2^+$ data, 
see \citeasnoun{jt236}.  
\citeasnoun{Hilico2000} further improved the accuracy of the energy computations 
to 10$^{-14}$ $E_{\rm h}$.  
There is an extensive literature on non-relativistic
H$_2^+$ calculations, but the references given provide a sufficiently
detailed picture for our purposes.  
The accuracy achieved at present
for the non-relativistic energy computations of H$_2^+$ and its deuterated analogs can be
considered as providing an ideal benchmark for studies on \hp\   and its D-analogs.

Extension of the electronic structure treatment to include relativistic and QED theory
\cite{Korobov06} reduced the discrepancy between calculations and
experiment \cite{PRLhdp07,RothPhysRev2006} to only about 100 kHz
($\approx 3 \times 10^{-6}$ \cm).  
\citeasnoun{Korobov06} supplemented
the relativistic and QED corrections, of the order of $\alpha^4$, where
$\alpha$ is the fine-structure constant, by corrections up to order $\alpha^6$.  
The resulting uncertainties in the low-lying energies are
about 300 MHz for the $\alpha^3$ and 2 MHz for the $\alpha^4$ terms.

Inclusion of $\alpha^3$ and $\alpha^4$ relativistic and QED
corrections in the first-principles determination of energies for
H$_2$ results in a discrepancy between theory and experiment for both
the low-lying levels and the dissociation energy, of only about 1 MHz
($\approx 3 \times 10^{-5}$ \cm) \cite{11KoPiLa.H2,14PaKoxx.H2,17PuKoPu.H2}.  
The contribution of QED terms to these energies
is given in the Supplementary Material of \citeasnoun{11KoPiLa.H2} and
its magnitude is up to 0.2 \cm.  
The latest computations include terms up to $\alpha^6$ \cite{16PuKoCz.H2}.

As mentioned before, the contribution of non-adiabatic effects in H$_2$
and H$_2^+$ can be computed extremely accurately \cite{09FaCzTa}.
However, most interesting for our purposes is the use of these
results for comparing the calculations of non-adiabatic effects
using the methods applicable also for polyatomic molecules, in
particular \hp.

\citeasnoun{80BuMoxx} developed perturbatively a
functional form for the non-adiabatic correction.  
This approximation can be expressed in the form of two different effective
nuclear masses, one for vibrations and another one for rotations.
These differ from the actual nuclear masses and mimic the non-adiabatic corrections.  
These masses may have a coordinate
dependence, in case of a diatomic molecule the distance between the two nuclei.  
\citeasnoun{96Moss} simplified this dependence and
represented both rotational and vibrational masses as constants.  
He 
(a) fixed the rotational mass correction to zero, that is the value of the
rotational mass is equal to the nuclear mass, and
(b) fitted the vibrational mass so that its use would reproduce the value of his
non-relativistic energies, calculated with an accuracy of 10$^{-5}$ \cm.  
This model was generalised to \hp\ by \citeasnoun{jt236}  as discussed extensively below.  
An alternative, empirical approach was suggested by \citeasnoun{03ScAlHia.H3+},
who introduced energy-dependent corrections to the band origins. 
This model was extended by \citeasnoun{alijah10} to higher energies and,
as also discussed below, has been tested for the \hp\ molecule.
When higher accuracy is necessary 
for the representation of non-adiabatic effects, 
the coordinate dependence of the rotational and vibrational
masses should be taken into account \cite{sl87,KutzJaquet}.  
The most recent calculation of non-adiabatic effects with a representation
in the form of coordinate dependent rotational and vibrational masses
is given by \citeasnoun{09PaKoxx.H2}.

\citeasnoun{13DiNiSa.H2} presented measurements and \ai\ computations
of the vibrational fundamentals of H$_2$, HD, and D$_2$ with an accuracy
of $2 \times 10^{-4}$ \cm, with a similar agreement obtained for the
dissociation energy \cite{16UbKoEi.H2}.  
For these studies the calculations were based on a fully correlated basis of 
exponential functions \cite{13PaYexx.H2} plus corrections for BO
\cite{14PaKoxx.H2}, relativistic, and QED effects. 
At this (high) level of accuracy, theory and experiment are in complete agreement.

There are other approaches to move beyond the usual BO approximation.  
One approach, the simultaneous consideration of all
electronic states, was explored by \citeasnoun{01Schwenke} and
\citeasnoun{09FaCzTa}.  
In particular, \citeasnoun{09FaCzTa} computed
energies for the three-body H$_2^+$ system using finite,
nuclear masses while maintaining the notion of a potential energy curve.  
Thus far this many coupled-states approach
has not yielded high-accuracy results, though it has improved our
understanding of non-adiabatic
treatments of molecules containing heavier nuclei \cite{Sch03b,jt282}.

An alternative approach is not to make the BO approximation and treat
the electron and nuclear motions simultaneously. 
A fully non-Born--Oppenheimer treatment of H$_2$ was recently presented by
\citeasnoun{17JoFoAd.H2}. 
They produced very accurate results, which
gave energy levels systematically 0.02 \cm\ above those presented by
\citeasnoun{11KoPiLa.H2}, who employed a more conventional approach
which makes an initial BO approximation, as discussed above. 
Of course, this shift means that the vibrational fundamental of H$_2$ is still
predicted with an accuracy similar to the measurement, but the 
dissociation energy is slightly underestimated.

\section{Electronic structure of \hp}
There is a long history of electronic structure computations on the
two-electron H$_3^+$ system. 
At the dawn of quantum chemistry, calculations by \citeasnoun{coulson_1935} demonstrated the 
unexpected stability if the ion and that it has an equilateral triangle equilibrium structure.
The first Born--Oppenheimer PES which gave results that
approached spectroscopic accuracy was due to \citeasnoun{mbb86} (MBB);
the MBB surface is not entirely \ai\ in that a single parameter was
tuned to improve the frequency of the $\nu_2$ bending fundamental.
Calculations using this surface played an important role in assigning
H$_3^+$ spectra, both in the laboratory \cite{jt83,jt102,jt125} and in space
\cite{jt80,jt98}. 
MBB used a full configuration interaction (FCI) method with a relatively large basis 
on a carefully designed grid of 69 points to define their PES.  
This grid became standard in many subsequent calculations.

\begin{table}
\begin{center}
\caption{Summary of \ai\ H$_3^+$ potential energy surfaces with
spectroscopic accuracy. The accuracy given is the difference
in the lowest energy point to that computed by \citeasnoun{ptf09}.}
\label{t:pes}

\begin{tabular}{llcrrll}
\hline \hline
\multicolumn{1}{c}{Authors}& 
\multicolumn{1}{c}{Method}&
\multicolumn{1}{c}{Dipole?}&
\multicolumn{1}{c}{$N_{\rm points}$}&
\multicolumn{1}{c}{Accuracy/cm$^{-1}$}& Designation\\
\hline 
\protect\citeasnoun{mbb86}       &Full CI     & yes & 69  & 160 & MBB  \\
\protect\citeasnoun{lf92}        &Hylleraas-CI& yes & 69  & 9   \\
\protect\citeasnoun{Rohse1994}   &CISD-R12    & no  & 69  & 1   \\
\protect\citeasnoun{Cencek1998}  &ECG         & no  & 69  & 0.04\\
\protect\citeasnoun{09BaCeJa.H3+}&ECG         & no  & 5900& 0.04\\
Pavanello {\it et al.} (2012)   &ECG         & yes &41655& 0.01& GLH3P \\
\hline  \hline
\end{tabular}
\end{center}
\mbox{}\\
\end{table}

Table~\ref{t:pes} charts the improvement in high-accuracy PES
computations starting with the MBB PES.  
Subsequent studies all
included explicit treatment of the electron-electron coordinate,
$r_{12}$, in the electronic structure calculation.  
As can be seen this leads to a dramatic improvement in the accuracy of the PES.  
\citeasnoun{92Anders.H3+}
used a Monte Carlo treatment to obtain, within a quoted uncertainty of
about 0.2 \cm, the precise electronic energy for H$_3^+$ at its equilibrium geometry.  
In practice this value has been superseded by
very extensive computations using explicitly correlated Gaussian (ECG) functions.  
The largest of these computations, due to
\citeasnoun{ptf09}, is used to benchmark the accuracy of the various
PES's considered in Table~\ref{t:pes}.\phantom{\citeasnoun{jt526}}

The model due to \citeasnoun{jt236} used the PES developed by
\citeasnoun{Cencek1998}, augmented by their relativistic correction
and an improved fit \cite{jt166} to their adiabatic corrections.
To allow for non-adiabatic effects, \citeasnoun{jt236} adapted the
approach of \citeasnoun{80BuMoxx}, who advocated the use of separate,
constant masses for the vibrational and rotational motions.
\citeasnoun{jt236} used the nuclear mass for the rotational motion and
effective vibrational masses based on the ones recommended for the
H$_2^+$ isotopologues by \citeasnoun{96Moss}.  
It should be noted that
the \citeasnoun{jt14} Hamiltonian used by Polyansky \&\ Tennyson (PT) is
formulated to exploit the cancellation between a vibrational and a
rotational term.  
Use of distinct masses for these two motions
therefore results in an extra, non-Born--Oppenheimer, term in the
Tennyson--Sutcliffe Hamiltonian.

\begin{table}
\begin{center}
\caption{Summary of less accurate \ai\ H$_3^+$ potential energy surfaces }    
\label{t:pes1} 

\begin{tabular}{llcrrll}
\hline \hline
\multicolumn{1}{c}{Authors}& 
\multicolumn{1}{c}{Method}&
\multicolumn{1}{c}{Dipole?}&
\multicolumn{1}{c}{$N_{\rm points}$}&
\multicolumn{1}{c}{Accuracy/cm$^{-1}$}\\
\hline 
\protect\citeasnoun{aguado:2000}  &Full CI     & no  & 8469& 20  &     \\
\protect\citeasnoun{11BaPrVi.H3+}        &DFT         & no  & 69  &315  \\
\protect\citeasnoun{VIE07:074309}   &Full CI         & no  & 8177 & 5   \\
\protect\citeasnoun{08VeLeAg.H3+}   &Full CI         & no  & 8469& 20  \\
This work   &Full CI, CBS     & yes &2500& 0.5 \\
\hline  \hline
\end{tabular}
\end{center}
\mbox{}\\
\end{table}

The PT computations reproduced the low-lying
rotation-vibration energy levels of H$_3^+$, H$_2$D$^+$, D$_2$H$^+$,
and D$_3^+$ to within a few hundredths of a \cm.  
Nevertheless, the accuracy
remains worse than that of computations based on the use of the best
semi-empirical spectroscopically determined PES \cite{jt175}, which
contained explicit allowance for adiabatic but not for non-adiabatic effects.
Furthermore, PT only considered low-lying
rotational states as their model makes no allowance for non-adiabatic
effects in the rotational motion, which should grow rapidly, as $J^4$, 
with rotational excitation, where $J$ is the quantum number
characterising the overall rotation.  
In what follows the model of PT is
used as a baseline against which more recent studies are compared.

For completeness, in Table~\ref{t:pes1} we also present a list of less
accurate electronic structure calculations of \hp\ PESs based on the
use of more conventional electronic structure methods, including
density functional theory (DFT). 
The reason for considering these calculations is twofold. 
First, these less accurate methods make it
inexpensive to compute many more points than the more accurate ones,
which resulted in purely \ai\ PESs, when more accurate methods
were still too expensive to be used to provide the coverage needed for
global calculations, see \citeasnoun{jt247} for an example. 
These calculations can potentially provide an ``unlimited'' number of points for
studies of global PESs if needed.  
Second, such calculations provide
the benchmark of the methods used when applied to systems with more
electrons, such as H$_5^+$ \cite{05XiBrBo.H5+,10AgBaPr.H5+}, for which the more sophisticated methods
listed in Table~\ref{t:pes} cannot be used.

To help understand the BO PES of H$_3^+$, for this study we computed 2500 full
configuration interaction (FCI) energy points, using the standard electronic
structure code MOLPRO \cite{MOLPRO}, with aug-cc-pV$n$Z ($n=5$ and 6) 
basis sets and produced a complete basis set (CBS) FCI surface, 
which differs from the ECG PES in absolute energy by less than 1 \cm.  
Nuclear motion calculations using this PES reproduce the
MARVEL energy levels of \hp\ with a standard deviation of 0.5 \cm,
which is only 5 times worse than with the most accurate PES currently available.

\section{Beyond the non-relativistic treatment}
Polyansky \&\ Tennyson employed the relativistic correction surface
computed by \citeasnoun{Cencek1998}.  
This correction is about 3 \cm\
but varies only weakly with internuclear separation, meaning that its
contribution to any calculated transition frequency is relatively minor.  
\citeasnoun{09BaCeJa.H3+} subsequently extended this surface,
which they computed as the expectation value of the complete
Breit--Pauli relativistic Hamiltonian using very accurate wave
functions based on ECGs.  
This relativistic surface, as far as we are
aware, meets the requirements for a high-accuracy calculation.  

As noted by \citeasnoun{jt581}, the smallness and smoothness of the
relativistic correction in H$_3^+$ is caused by the almost complete
cancellation between the two most important first-order corrections,
the one-electron mass-velocity and Darwin terms,
together usually denoted as MVD1 \cite{01TaCsKl.methods}.  
This cancellation of contributions results in another interesting effect.  
Superficially the QED contribution to the
energy levels should be much less than the relativistic contribution,
as QED effects are generally about 5 \% of the relativistic effect.
However, as QED is 5 \% of only one part of the scalar relativistic effect --
that is of the one-electron Darwin term \cite{jt265}, 
in the case of H$_3^+$ the overall contribution of QED
to the energy levels is comparable to the overall relativistic (MVD1) one. 
This means that it is necessary to allow for the QED effects in
all accurate calculations of the \hp\ spectrum.

The relativistic surfaces used by \citeasnoun{jt526} and \citeasnoun{jt535} 
were limited to 30 000 \cm, as all experimental energy levels with which
comparisons could be made lie well below this energy threshold. 
For the purpose of making the ECG-based global $\rm H_3^+$ potential 
(named GLH3P by \citeasnoun{jt526}) a fully global potential, a fit of the relativistic 
energies to a  global set of geometries is mandatory. 
We produced such a fit as part of this study
and the resulting analytic surface reproduces the  set of \ai\ points with
a standard deviation of about 0.02 \cm. Geometries with energies from
0 to 43~000 \cm\ are used in the fit and the resulting global
relativistic surface contains 40 constants. 
This surface is given in the Supplementary Material. An outstanding issue is
the stability of this and other surfaces over the entire range of coordinates.
A high accuracy surface that satisfies this criterion which is essential for
studies of spectra above dissociation as well as reaction dynamics is 
currently under construction.

Consideration of the effects introduced by
QED can be important for the accurate
prediction of vibration-rotation spectra of even H-containing molecules.
\citeasnoun{jt581} used the methodology of \citeasnoun{jt265} to compute QED
corrections to the spectrum of H$_3^+$. 
Lodi {\it et al.} found that
including QED effects leads to shifts of about 0.1 \cm\ in the predicted
vibrational band origins but combining them with the Polyansky \&\ Tennyson
model actually made the results worse by roughly this amount. 
This was the first indication that the excellent results of PT may have 
been, at least partially, fortuitous.

\section{Fitting the potential energy surface}
There are a number of
global surfaces available for the ground electronic state of the H$_3^+$ ion
\cite{jt55,jt65,jt202,jt247,aguado:2000,VIE07:074309,08VeLeAg.H3+,jt526}. 
As can be seen from Table~\ref{t:pes}, 
a number of high-accuracy PESs were based
on the 69 point grid originally designed by \citeasnoun{mbb86}. 
Clearly, when constructing global 
surfaces from \ai\ data a much more extensive grid is required.
 
\citeasnoun{jt526} computed a high-accuracy PES with over 40~000 points; 
this allowed them to 
both produce a global surface and test the coverage of MBB's 69-point grid.
\citeasnoun{jt535} compared the multipoint GLH3P fit of  \citeasnoun{jt526} 
with a fit to just the standard 69 points. 
This gave an interesting result: while differences for some of the vibrational levels 
were found to be very small, there are differences of a few tenths of a \cm\   for 
levels up to the barrier to linearity and between a few
to tens of \cm\   for the levels above it. 
This demonstrates that MBB's 69
points are insufficient to accurately characterise the PES of \hp. 

In order to understand better the influence of the number of points
and the density of the grid on the final PES, we computed several PESs
using 69 MBB points, then the same 69 points plus 300 points and so
forth up to the full GLH3P grid. 
These calculations, 
summarised in Table~\ref{t:fit}, show that the use of additional points
in the PES fit has a significant effect on the computed vibrational energies. 
In particular, moving from 69 points to a much larger set
gradually {\it increases} the  observed minus calculated residues for 
vibrational term values
lying below 7000 \cm\ from about 0.05 \cm\ to about 0.1 \cm.  
However, levels above 7000 \cm\ are improved in
comparison with calculations using the \citeasnoun{jt236} surface, and
result finally in the accuracy of GLH3P, demonstrated by
\citeasnoun{jt526} and \citeasnoun{jt535} to be about 0.1 \cm\ up to 17~000 \cm. 
These calculations demonstrate that the extremely
high accuracy of the \citeasnoun{jt236} calculations for the then
available levels, which all lie below 7000 \cm, was indeed fortuitous.

\begin{table}
\begin{center}
\caption{Comparison of calculated vibrational term values for 
several different PESs. Computed energies, in \cm, are given as observed minus calculated, and they are   compared
to the empirical MARVEL energies of \citeasnoun{13FuSzMa.H3+}. All calculations
used relativistic, QED and DBOC corrections plus the PT non-adiabatic model.}
\label{t:fit}
\begin{tabular}{crrrrrr}
 \hline\hline
$v_1v_2\ell$ & MARVEL   &   GLH3P& BO-69 & BO-2500& BO-69&       BO-369\\  
             & Expt.    &   ECG &  Full CI & Full CI & ECG & ECG\\
\hline                                    
0 1 1&   2521.408   &   0.11&    0.17 &       0.11  &      0.08 & 0.12\\
0 2 2&   4998.048   &   0.15&    0.16 &       0.12  &      0.08 & 0.16 \\
1 1 1&   5554.061   &$-$0.14&    0.19 &    $-$0.14  &   $-$0.03 &$-$0.09 \\
0 3 3&   7492.911   &   0.13& $-$0.03 &    $-$0.03  &   $-$0.07 &0.13  \\
2 2 2&   10645.377  &   0.06&    1.51 &    $-$0.76  &     3.14  &0.06 \\
0 5 1&   10862.901  &   0.15&    1.95 &    $-$1.00  &     1.33  &0.20\\
3 1 1&   11323.096  &$-$0.03&    1.42 &    $-$1.46  &     3.05  &0.01\\
0 5 5&   11658.397  &   0.08&    1.45 &    $-$1.39  &     3.47  & 0.08\\
2 3 1&   12303.363  &   0.02&    0.78 &    $-$2.03  &     2.77  & $-$0.06\\
0 6 2&   12477.378  &$-$0.02&    1.18 &    $-$2.41  &     2.07  & $-$0.04\\
0 7 1&   13702.372  &$-$0.22&    1.65 &    $-$5.21  &     2.05  & $-$0.26\\
0 8 2&   15122.801  &   0.15&    2.39 &    $-$5.59  &     7.35  &  0.00\\
\hline  \hline
\end{tabular}
\end{center}
\end{table}

\section{Nuclear motion computations}
Accurate solutions of the nuclear motion problem for low-lying states
of H$_3^+$ and its isotopologues have been obtained by a number of groups
\cite{jt181,jt236,ALI95:1105,ALI95:1125,96AlBe,jaquet:2002,03ScAlHia.H3+,03ScAlHib.H3+,08VeLeAg.H3+,09BaCeJa.H3+,alijah10,JAQ10:157,13FuSzFaCs,14MaSzCs.H3+,jt512}.
These extensive studies confirm that within the BO approximation different
approaches to the variational solution of the nuclear motion problem
all yield essentially the same answers and that the numerical
uncertainties introduced at this stage of the calculation are negligible.  
This means that for studying
spectra of H$_3^+$ the principal issue for the accurate determination of
rovibrational states is the method used to treat the breakdown of the BO approximation. 
This aspect of \hp\  spectroscopy is considered in detail
in section 9.

The (nearly) complete set of bound vibrational levels of H$_3^{+}$ and
its deuterated isotopologues have also been determined several times
using accurate PESs \cite{jt358,jt370,jt387,jt389,10SzCsCz.H3+}.  
Such studies require a PES which has correct dissociative behavior, a
property which few fitted PESs possess.  
The so-called PPKT2 PES \cite{jt247,jt370} satisfies this criterion.  
A few related relevant results are as follows: 
(a) the lowest dissociation energy ($D_0$) of \hp,
corresponding to the reaction H$_3^+$ $\rightarrow$ H$_2$ + H$^+$, is
34~912 \cm\   \cite{jt370}; 
(b) below $D_0$ there are at least 688
even-parity and 599 odd-parity vibrational states, as computed by
\citeasnoun{10SzCsCz.H3+}. 
Computing the last few states below the
first dissociation limit is rather problematic for all molecules.
This is partly due to the fact that rovibrational Hamiltonians
expressed in internal coordinates must have singular terms in their
kinetic energy part, requiring a careful choice of basis functions in
variational and near-variational treatments.
Furthermore, the last
few states extend to very large values along the dissociation coordinate,
requiring the use of extended basis sets \cite{jt358}.  
The long-range nature of the
H$_3^+$ potential results in the system supporting a number
of asymptotic vibrational states \cite{jt358}. Trial numerical
studies suggested that these states are not that sensitive to the
precise form of the long-range potential used \cite{jt389}.
However, there is definely more work to be done on this problem.

\section{Born--Oppenheimer breakdown} 
The mass-dependent non-BO effects can be separated into adiabatic,
or diagonal Born--Oppenheimer correction (DBOC) \cite{hys86,Kut97}, 
and non-adiabatic effects.
 
The first adiabatic surface was determined by \citeasnoun{jt147} by
inverting experimental data. 
More recently \ai\ techniques have been used for the same purpose. 
In particular,
the adiabatic  surface used by \citeasnoun{jt526} and \citeasnoun{jt535} 
for their GLH3P calculations was produced (as in the case of the relativistic surface)
up to 30 000 \cm, as all the experimentally known energy levels were much below this value. 
The surface was obtained by the fitting of 3300 
\ai\ points with a standard deviation of 0.017 \cm\ employing 98 parameters. 
In order to make the GLH3P surface genuinely global, 
we fitted 5500 points (out of 6000 available) for 
geometries with energies from 0 to 39 000 \cm. 
The number of parameters of the surface is 90 and the standard deviation 
of the fit is 0.22 \cm. 
This fit is an order of magnitude worse than that of the above mentioned more limited surface.
However, the accuracy is comparable to the accuracy of the BO surface and 
provides a good basis for global nuclear motion calculations.
This surface is given in the Supplementary Material, as is a spreadsheet
illustrating the effect of correctly treating the adiabatic correction
at high energies.
 
Next, let us consider the non-adiabatic effects on the nuclear motions of \hp. 
Non-adiabatic effects constitute the weak point of all nuclear motion calculations on \hp. 
As a  result  the
remaining 0.1 \cm\ residues in the computed \hp\ ro-vibrational energy levels
are mainly due the solution
of the non-adiabatic problem. 

\citeasnoun{06AlHi} provided a thorough analysis of the non-adiabatic
effect for the rovibrational levels of H$_3^+$ based on some simple
linear and quadratic correction functions.  
This study was hindered by
the fact that basically no accurate experimental rovibrational
energies higher than about 10~000 \cm\ were available in 2006.  
After the availability of experimental (MARVEL) energy levels,
\citeasnoun{14MaSzCs.H3+} compared the ultimate Born--Oppenheimer
rovibrational energies computed utilizing the adiabatic GLH3P PES with
the highly accurate MARVEL levels.  
The exceptional quality of the
GLH3P adiabatic PES, namely that it reproduces all the known term
values when employing nuclear masses for both rotational and
vibrational motion with an RMS error of just 0.19 \cm, is clear from
this comparison. 

The model based on constant but motion-dependent masses 
(different vibrational and rotational masses)
provides an appealing choice to represent non-adiabatic effects, as it  
(a) is conceptually 
simple; (b) keeps the notion of a PES almost intact (see below); and
(c) has been proved to improve computed energies with respect to
experimental values, in the case of \hp\ by a factor of two
\cite{14MaSzCs.H3+}.  The theoretical basis for such a model has been
devised for diatomics by \citeasnoun{Herman1966} and by
\citeasnoun{bm77}, who derived effective Hamiltonians
incorporating, in the absence of avoided crossings, most of the
non-adiabatic effects.  Their treatment yielded separate,
coordinate-dependent masses for vibration and rotation, and a
correction term to the potential, which demonstrates that the effect
of non-adiabatic coupling cannot be described solely by adjusting the
PES.  Introducing different vibrational and rotational masses, but
keeping them constant, may be seen as a lowest-order approximation.

The next step would be the
introduction of coordinate-dependent mass surfaces (CDMS) for both the
rotational and vibrational masses.  
\citeasnoun{14MaSzCs.H3+} considered the form of the Hamiltonian 
when the masses are allowed to have coordinate dependence.  
They showed that, at least within the fully numerical and black-box-like GENIUSH code
\cite{09MaCzCs.method,11FaMaCs}, both the motion-dependent and the
CDMS models can be implemented with relative ease.  
This opens the route toward a systematic improvement of the theoretical description
of second-order non-adiabatic effects in polyatomic molecular systems.
Another principal and far-reaching conclusion of this study is that even in the case of
constant but motion-dependent masses the computed rovibrational energy
levels depend on the embedding of the body-fixed frame utilized for the computation.  
Among the embeddings tested it was only the
\citeasnoun{35Eckart.method} embedding with a symmetric triangular reference
structure which remained invariant under the permutation of the protons.  
Except for this case of a permutationally invariant
embedding, an artificial splitting characterizes the computed
degenerate rovibrational eigenvalues.  
This splitting increases with $J$ and
exceeds the assumed accuracy of the computation by about $J=5$.

When optimizing the vibrational mass using 15 selected measured
high-accuracy low-energy transitions, see Table I of
\citeasnoun{14MaSzCs.H3+}, while keeping the rotational mass constant
at the nuclear mass of H, the optimal vibrational mass of the proton
turned out to be higher by about one-third of an electron mass than the nuclear mass.  
This value is substantially less then the optimal mass given by \citeasnoun{96Moss}
for H$_2^+$ as used in the PT model, where the correction is almost half
of the mass of an electron.  
This result is extremely similar to the
conclusion of a recent work of \citeasnoun{jt566}, who obtained a
core-mass surface from a simple Mulliken population analysis carried
out for H$_3^+$.  
Using this mass surface the authors determined
effective masses for each vibrational state in an iterative procedure,
and they obtained extremely similar mass corrections for the $00^{0}$
and $01^{1}$ states.  
Using the optimized vibrational mass of
\citeasnoun{14MaSzCs.H3+}, the RMS deviation between the
``non-adiabatic'' first-principles and the 15 empirical (MARVEL)
rovibrational energies becomes only 0.008 \cm. 
This is down from an
RMS discrepancy of 0.19 and 0.10 \cm\ employing the nuclear and the Moss masses, respectively. 
The improved differences basically
correspond to the internal accuracy of the experimental MARVEL energy levels of H$_3^+$.


An \ai\ study by
\citeasnoun{15AlFrTy.H3+} of the non-adiabatic coupling terms  
showed that up to four electronic states
need to be included in conventional non-adiabatic dynamical
calculations, depending on the accessible nuclear configurations and energy. 
This demonstrates that the use of motion-dependent masses is
the most promising way for further improvement of the computed bound energy levels.

When discussing high accuracy, first principles treatments of H$_2^+$ and H$_2$ we noted that 
computations performed without making the BO approximation gave results competitive with
the one based on more conventional treatments augmented by BO breakdown corrections. 
This is not true for H$_3^+$, as it is 
currently not possible to perform accurate fully non-BO treatments of triatomic species
(L. Adamowicz, private communication, 2013; E. M\'{a}tyus, private communication, 2016), 
even though such treatments are possible on five-particle systems with only
two heavy particles (atoms) \cite{09StBuAd}.

To conclude this section, let us note that the \dpp\ ion is much less
well studied experimentally than \hp.  Nevertheless, as the
non-adiabatic effects are mass-dependent, the influence of them on the
energy levels of \dpp\ is significantly smaller.  While there is more
information on the levels of H$_2$D$^+$ and D$_2$H$^+$, the
lower-symmetry of these mixed isotopologues introduces extra,
symmetry-breaking non-BO terms \cite{jt175} making such a comparison
much less straightforward.  With an extensive set of \dpp\
experimental energy levels, we could more reliably demonstrate that
the remaining 0.1 \cm\ residues in \hp\ energy level calculations are
due to non-adiabatic effects and, in other words, that all other
components of the calculations are accurate to 0.01 \cm\ or better.
This would put \ai\ calculations on the \hp\ molecular ion in their
expected position between H$_2$, with an accuracy of $10^{-4}$ \cm,
and water \cite{jt550} and H$_2$F$^+$ \cite{h2fp}, with an accuracy of
0.1 \cm\ for \ai\ calculations.

\section{Transition intensities}
Use of H$_3^+$ spectra for remote sensing, for example during astrophysical
applications, relies on transition intensities which in turn establish column densities. 
However, laboratory experiments rarely prepare H$_3^+$ ions
in thermal equilibrium; a modern exception to this is the cold
populations prepared using an ion-trap apparatus \cite{jt413}. 
This means that absolute laboratory measurements of transition intensities are unusual. 
Indeed, \citeasnoun{98McWa} provided a rare example of an H$_3^+$
spectrum with absolute intensities; there are a number of other
examples where transitions from the same lower state have been measured
under the experimental conditions to give accurate relative
intensities \cite{jt289,jt413,jt587}.

In the absence of measured line intensities, theory has taken on the
role of providing line intensities to aid modeling and remote sensing
of H$_3^+$ ions.  These are largely provided in the form of line lists
based on either computed transition frequencies \cite{jt169,jt478} or
empirical ones where available \cite{jt107,jt666}.  Experience for a
number of molecules suggests that accurate intensities are best
computed using \ai\ dipole moment surfaces (DMS) \cite{jt156,jt573}.
Computing transition intensities therefore requires accurate wave
functions and a high quality \ai\ DMS.  As shown in Table~\ref{t:pes},
there are a number of DMSs available although not all high accuracy
PES computations have been extended to provide a DMS.

Although the non-BO contribution to the transition intensity can be
expected to be small \cite{09HoVaEd.DMS}, the DMS for the asymmetric
isotopologues H$_2$D$^+$ and D$_2$H$^+$ is significantly different
from that for H$_3^+$. This is a result of the separation of the
centre-of-charge and centre-of-mass in the asymmetric systems which
leads to a permanent dipole moments of approximately 0.60 D and 0.46 D
for  H$_2$D$^+$ and D$_2$H$^+$, respectively. Again there are no
direct measurements of these dipoles or any associated 
transition intensities.

In the absence of high accuracy, absolute experimental intensity data it is
difficult to be definitive about the accuracy of the available DMSs and the
associated transition intensities. 
The most stringent test is currently
provided by the relative intensity measurements of \citeasnoun{jt587},
who probed 18 lines starting from states with $J=0$ and 1 at visible wavelengths. 
These results show that even the relatively simple DMS of
\citeasnoun{lf92} gives very good results. 
This DMS is based on
calculations performed at the 69 grid points of MBB and a modest fit employing
only 7 constants.  
The much more extensive grid of dipole moments
presented by \citeasnoun{jt587} should allow a more extended and
possibly more accurate DMS to be produced. 
Use of this extended grid
of points raises issues with fitting similar to those encountered with
PES fits and discussed above.

\section{Behaviour at dissociation}
While it is straightforward to compute rovibrational states of \hp\ and
its deuterated isotopologues lying below about 20~000 \cm, or halfway
to dissociation, this is not so for high-lying states.  
Studies have probed the highest bound rotational states of the H$_3^+$ system
\cite{jt64,13JaCaxx.H3+,13Jacquet.H3+} predicting that the highest
bound state has $J=46$. 

In order to study the near dissociation spectrum of H$_3^+$ observed
by Carrington and co-workers, it is necessary to consider
vibrationally excited states both just below and just above dissociation.  
As discussed above, a series of studies have been
performed focused on representing all the vibrational states up to dissociation
\cite{jt100,jt132,jt370,10SzCsCz.H3+}.  
Of course it was above dissociation that Carrington and co-workers
observed their very complex and structured spectrum. 
Fully quantal studies approaching the accuracy of experiments
in this region are extremely challenging. \citeasnoun{97MaTaxx.H3+}
performed very early calculations identifying vibrational (Feshbach) resonances. 
However, at that stage there were no realistic global PESs available. 
More recent studies identifying resonances
\cite{jt443} have been based on more realistic
surfaces and have begun to consider the role of rotational motion
which gives the shape resonances that are the key to understanding the 
near-dissociation spectrum \cite{88LlPoxx.H3+,88ChChxx.H3+}.
There is clearly a need to do more theoretical work on this problem.
As is clear from reviewing the available models, methods, and codes, these future
studies would need to combine a highly accurate global PES, including even QED
correction, and a good representation of first- and second-order adiabatic
corrections, perhaps in the form of coordinate-dependent mass surfaces.
Work in this direction is currently being undertaken \cite{jtgh3p}.
Furthermore, efficient computation and identification of rovibrational resonance
states, including the reliable computation of their intensities, is needed, an
area under constant development \cite{jt253,jt494,11Moisey.book,13SzCsxx.H2O}.

\section{Conclusions}
\hp\  is an astronomically important molecular ion which also
provides a key benchmark of theoretical treatments for polyatomic molecules.
{\it Ab initio} treatments of H$_3^+$ remain many orders of magnitude less
accurate than those available for diatomic hydrogenic systems (H$_2^+$ and H$_2$). 
The main unresolved problem for high accuracy predictions of the
rotation-vibration spectrum of H$_3^+$ and its isotopologues is the
treatment of non-adiabatic effects beyond the Born--Oppenheimer approximation.
A number of methods of including non-adiabatic effects have been explored;
the most promising appears to be the use of coordinate-dependent effective
masses.

While vibration-rotation spectra involving low-lying states of \hp\ is well understood, 
the same cannot be said of its near-dissociation spectrum.
Experimental photodissociation spectra of \hp\ and its isotopologues recorded
three decades ago provide a window into the very complex structure of the 
nuclear motion states both just below and just above the lowest dissociation limit. 
This problem still awaits a proper theoretical elucidation.

\section*{Acknowledgements}
We thank members of the ExoMol team and Steven Miller for many helpful discussions.  
Some of the work reported was supported by the ERC under the
Advanced Investigator Project 267219. 
This work was partly supported by Russian Fund for Basic Research.
Collaboration between the UCL and ELTE groups has greatly benefited from
the support of the  COST action MOLIM: Molecules in Motion (CM1405).
AGC thanks the support of NKFIH (grant no. K119658).

\section*{References}
\bibliographystyle{jphysicsB}

\end{document}